\documentclass[conference, a4paper]{IEEEtran}
\IEEEoverridecommandlockouts
\usepackage{cite}
\usepackage{amsmath,amssymb,amsfonts}
\usepackage{algorithmic}
\usepackage{graphicx, dblfloatfix}
\usepackage{sidecap}
\usepackage{subcaption}
\usepackage{url}
\usepackage[super]{nth}
\DeclareGraphicsExtensions{.png, .pdf}
\usepackage{tikz}
\usetikzlibrary{positioning}

\usepackage{textcomp}
\usepackage{xcolor}
\def\BibTeX{{\rm B\kern-.05em{\sc i\kern-.025em b}\kern-.08em
    T\kern-.1667em\lower.7ex\hbox{E}\kern-.125emX}}
\graphicspath{{Figure/}}
\begin{document}

\title{Towards a Perceived Audiovisual Quality Model for Immersive Content\\
}

\author{\IEEEauthorblockN{Randy Frans Fela}
\IEEEauthorblockA{\textit{SenseLab} \\
\textit{FORCE Technology}\\
Hørsholm, Denmark \\
rff@force.dk}
\and
\IEEEauthorblockN{Nick Zacharov}
\IEEEauthorblockA{\textit{SenseLab} \\
\textit{FORCE Technology}\\
Hørsholm, Denmark \\
nvz@force.dk}
\and
\IEEEauthorblockN{Søren Forchhammer}
\IEEEauthorblockA{\textit{Dept. Photonics Engineering} \\
\textit{Technical University of Denmark}\\
Kgs. Lyngby, Denmark \\
sofo@fotonik.dtu.dk}
}
\IEEEpubid{\makebox[\columnwidth]{978-1-7281-5965-2/20/\$31.00 \copyright 2020 IEEE \hfill} \hspace{\columnsep}\makebox[\columnwidth]{ }}
\maketitle

\begin{abstract}
This paper studies the quality of multimedia content focusing on 360 video and ambisonic spatial audio reproduced using a head-mounted display and a multichannel loudspeaker setup. Encoding parameters following basic video quality test conditions for 360 videos were selected and a low-bitrate codec was used for the audio encoder. Three subjective experiments were performed for the audio, video, and audiovisual respectively. Peak signal-to-noise ratio (PSNR) and its variants for 360 videos were computed to obtain objective quality metrics and subsequently correlated with the subjective video scores. This study shows that a Cross-Format SPSNR-NN has a slightly higher linear and monotonic correlation over all video sequences. Based on the audiovisual model, a power model shows a highest correlation between test data and predicted scores. We concluded that to enable the development of superior predictive model, a high quality, critical, synchronized audiovisual database is required. Furthermore, comprehensive assessor training may be beneficial prior to the testing to improve the assessors' discrimination ability particularly with respect to multichannel audio reproduction. 

In order to further improve the performance of audiovisual quality models for immersive content, in addition to developing broader and critical audiovisual databases, the subjective testing methodology needs to be evolved to provide greater resolution and robustness. 
\end{abstract}

\begin{tikzpicture}[overlay, remember picture]
\path (current page.north) node (anchor) {};
\node [below=of anchor] {%
2020 Twelfth International Conference on Quality of Multimedia Experience (QoMEX)};
\end{tikzpicture}

\section{Introduction}
Since the introduction of virtual reality, 360 video has recently become a popular way of presenting immersive content and due to its potential applications, numerous efforts can be found for improving and assessing its quality using objective and subjective measures \cite{li2019state, perez2019miro360, schatz2017towards, xu2018assessing}. Objective measures of 360 video aims to quantify the quality the reconstructed video based on its distortion compared to the original video in the form of a peak signal-to-noise ratio (PSNR) metric. Due to the various projection formats of 360 video from spherical to 2-dimensional plane (e.g. equirectangular (ERP), cubemap (CMP), etc.) and vice versa, this opens the possibility to calculate PSNR-variant metrics as proposed in \cite{sun2016ahg8, he2016ahg8, yu2015framework, zakharchenko2016ahg8}. When considering the user experience, users often only view a proportion of spherical projection of 360 video, where a dynamic viewport-based PSNR metric has been proposed to represent the user behavior throughout the video. 360 video processing workflow and a study of its PSNR related metrics have been shown in earlier studies and JVET-J1012 for common test condition and evaluation procedures \cite{li2019state, hanhart2018360, boyce2017jvet}. The processing workflow enables PSNR metrics in 360 video to be measured in three phases,  namely codec, cross-format spherical and end-to-end spherical metrics. Similarly, subjective evaluations have been conducted earlier which generated the distorted videos using identical coding parameters of quantization parameters (QP) and resolution \cite{hanhart2018360, tran2017subjective}. These findings are in agreement with our own that the perceived quality is proportional to resolution and inversely proportional to QP. 

In recent years, the trend of 360 video applications in multimedia platforms (i.e. Youtube, Facebook, and Google) has been paired with spatial audio formats such as  ambisonics which allows 3D auditory sensation while watching omnidirectional video. A low-bitrate audio codec plays an important role in streaming applications and therefore with this has come a growing interest in evaluating the quality of compressed ambisonic scenes. The findings from \cite{rudzki2019perceptual} stated that lower bitrate ambisonic has a lower quality score and higher localization error.

The use of 360 videos with ambisonics has been demonstrated in order to assess specific perceptual attributes \cite{kentgens2018spatial, olko2017identification}. However, to the best of our knowledge, a study on perceived audiovisual quality and its interaction is still relatively unexplored. This paper describes a preliminary study investigating the perceptual quality of audio, video, and audiovisual quality of 360 videos and ambisonic reproduction. The aim of this study is to highlight the potential applications and limitations in this area of interest including an initial understanding of audiovisual interaction towards the future multimodal audiovisual quality models. This work addresses the following questions:

\begin{table*}[ht]
\caption{Characteristics of the audio video contents}
\begin{center}
\begin{tabular}{|l|c|c|c|c|c|c|}
\hline
\textbf{Video}&\multicolumn{3}{|c|}{\textbf{Video Characteristics}}&\multicolumn{3}{|c|}{\textbf{Audio Characteristics}} \\
\cline{2-7} 
\textbf{} & \textbf{\textit{Lighting condition}}& \textbf{\textit{Motion activity}}& \textbf{\textit{Spatial complexity}}& \textbf{\textit{Type}}& \textbf{\textit{Character}}& \textbf{\textit{Source}} \\
\hline
1. Duomo& Low light& Low& Simple& Orchestra, solo& Reverb, clarity& Static \\
\hline
2. MotoCani& Bright, daylight& Fast& Medium& Natural, mechanical, human& Low-freq dominant& Dynamic\\
\hline
3. Autodafe& Low light& Low& Simple& Orchestra, group singer& Reverb& Static, dynamic \\
\hline
14. ParcoDucale& Bright, daylight& Low& Medium& Natural, human& Ambient dominant& Static, dynamic\\
\hline
5. Fisarmonica& Bright, daylight& Fast& Complex& Music, mechanical& Reverb-ambient& Dynamic\\
\hline
\end{tabular}
\label{tab1}
\end{center}
\end{table*}

\begin{figure*}[ht]
\centering
\begin{subfigure}{0.18\linewidth}
    \includegraphics[width=\linewidth]{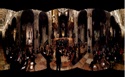}
    \subcaption{Video 1}
    \label{fig:Cathedral}
\end{subfigure}
\begin{subfigure}{0.18\linewidth}
    \includegraphics[width=\linewidth]{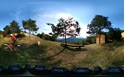}
    \subcaption{Video 2}
    \label{fig:Motocycle}
\end{subfigure}
\begin{subfigure}{0.18\linewidth}
    \includegraphics[width=\linewidth]{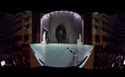}
    \subcaption{Video 3}
    \label{fig:Opera}
\end{subfigure}
\begin{subfigure}{0.18\linewidth}
    \includegraphics[width=\linewidth]{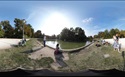}
    \subcaption{Video 4}
    \label{fig:Park}
\end{subfigure}
\begin{subfigure}{0.18\linewidth}
    \includegraphics[width=\linewidth]{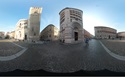}
    \subcaption{Video 5}
    \label{fig:Piazza}
\end{subfigure}

\caption{Equirectangular view of the testing videos.}
\label{fig}
\end{figure*}

\begin{itemize}
\item What are the effects of the video encoding parameters on video quality score?
\item What are the effects of number of audio channels and bitrates on audio quality score?
\item How do perceived audio and video quality relate and combine to perceived audiovisual quality?
\item What is an appropriate testing methodology to study these characteristics?
\end{itemize}

Based on these questions, we established experiments based on common approaches used in audio and video quality evaluation. Due to the  large number of experimental parameters for full factorial design, we introduced optimal custom design for audiovisual quality assessment in a manageable trial size.

\section{Study Description and Experiments}
\subsection{Stimuli}\label{AA}
Audiovisual stimuli were provided by the Jump Video Dataset from the University of Parma \cite{noauthor_index_nodate}. This source contains 360 video with ambisonic spatial audio. Five 360 videos ($\sim$35s) were carefully selected covering different audiovisual context as described in Table I. Snapshots of content can be seen in Fig. 1. The footage was recorded using a 32-capsule spherical EigenMike microphone array and circular array of 8 GoPro Session 4 cameras. After post-processing, the output of each video was a 4K resolution (3840x1920) equirectangular projection (ERP) format with 8-bit depth, 29.97 frame rates, $\sim$30 Mbps bitrates in YUV 4:2:0 color space and chroma subsampling. The raw 32-channels A-format audio were post-processed into Ambix FOA (4 channels) B-format, PCM sampling format and temporally aligned with the video. The audio sampling rate was 48 kHz with the 16 bits and the total bitrates of 3.072 Mbps (768 kbps/channel).

\subsection{Encoding and Decoding}
The video material was encoded using FFmpeg (with libx264) with frame rate 29.97 fps in a GOP structure of “IBBP” with a GOP size of 16. Twenty encoding settings were applied to create quality degradation for each video corresponding to combination of the original quality and four quantization parameter (QP) values of 22, 27, 32 and 37 and four resolutions of 3840x1920 pixels (4K), 2560x1280 pixels (2.5K), 1920x1080 pixels (fHD) and 1280x720 pixels (HD). In total, 100 videos were used for this study.

Lossy audio coding was performed in FFmpeg with low complexity Advanced Audio Coding (AAC-LC) format to generate three low-bitrate ambisonics files in 64 kbps, 128 kbps and 256 kbps. Note that the bitrate is the total for 4 channels (4 ambisonic channels). Uncompressed clips were also included into the test. Adobe Audition CC 2019 with Spatial Audio Real-Time Application (SPARTA) VST plug-ins installed \cite{mccormack2019sparta} was employed to decode each audio clips into respective channels of 5.0, 11.0 and 22.0 loudspeaker setup, according to ITU-R BS.2051, generating 60 audio streams.

\subsection{Measures}
Objective video quality metrics including PSNR and its variants were calculated using HM16.16 reference software and 360Lib software package \cite{boyce2017jvet}. The quality measure of a video stream is the average of its frame quality values (I-P-B frames and YUV 4:2:0 color format). Based on the PSNR processing chain for 360 video quality measures, these PSNRs cover three different phases namely codec PSNRs, Cross-Format (CF) PSNRs and End-to-End (EE) PSNRs \cite{hanhart2018360}.

\begin{figure*}[h]
\centering
\begin{subfigure}{0.28\linewidth}
    \includegraphics[width=\linewidth]{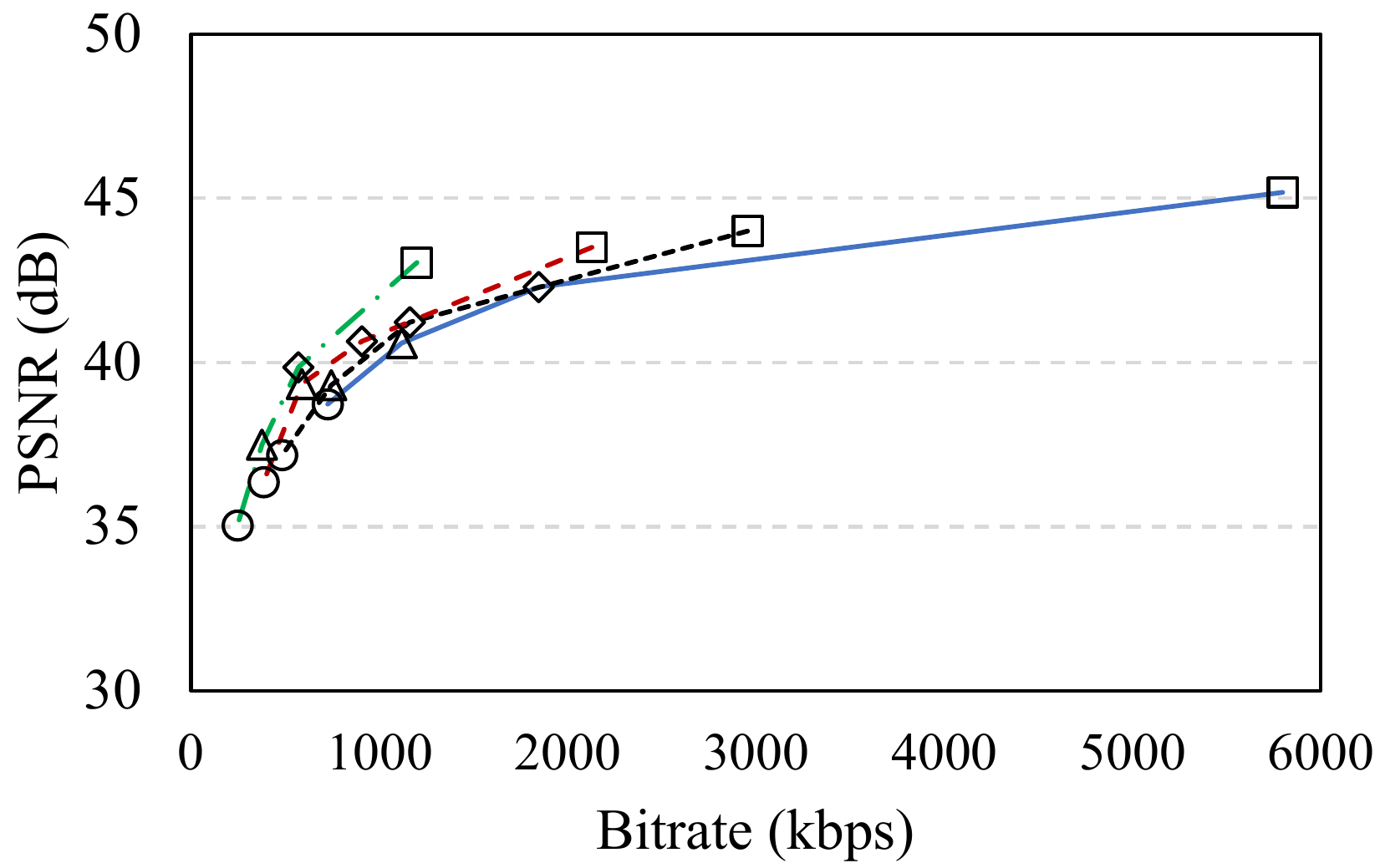}
    \caption{Video 1}
\end{subfigure}
\begin{subfigure}{0.28\linewidth}
    \includegraphics[width=\linewidth]{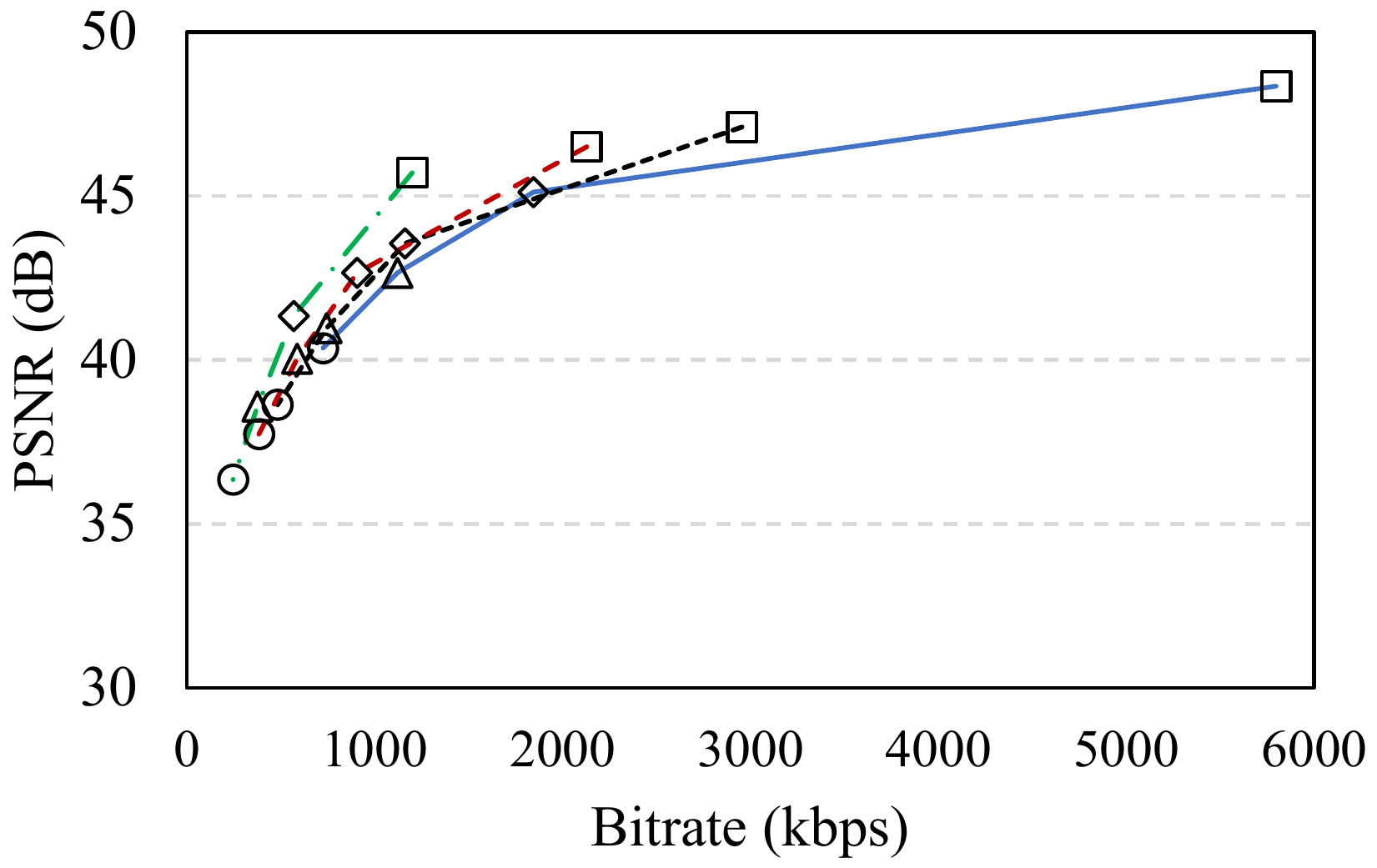}
    \caption{Video 3}
\end{subfigure}
\begin{subfigure}{0.365\linewidth}
    \includegraphics[width=\linewidth]{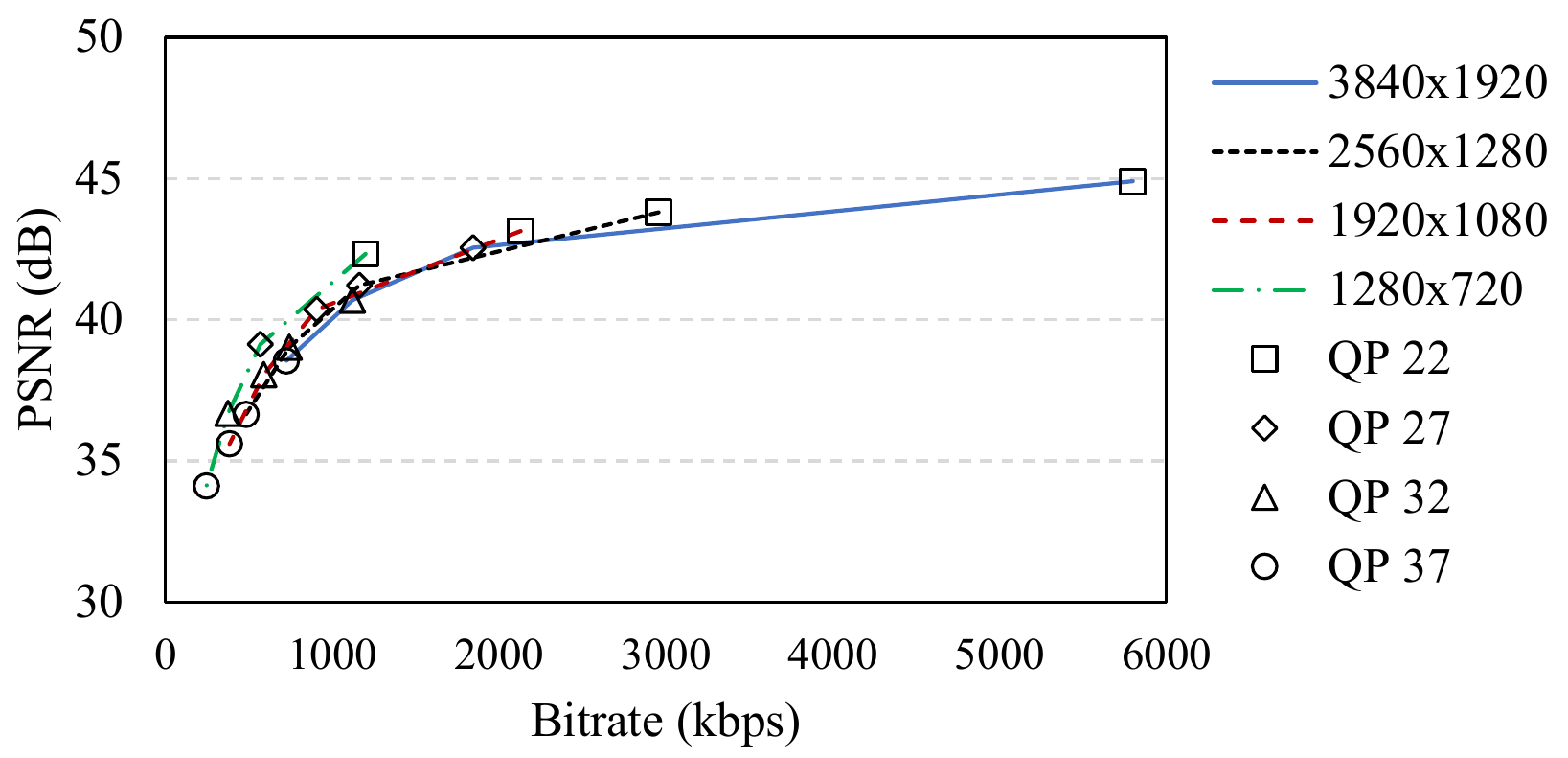}
    \caption{Video 5}
\end{subfigure}
\caption{Codec Peak signal-to-noise ratio (PSNR) vs video bitrate.}
\end{figure*}

\begin{figure*}[h]
    \centering
    \begin{subfigure}{0.30\linewidth}
    \includegraphics[width=\linewidth]{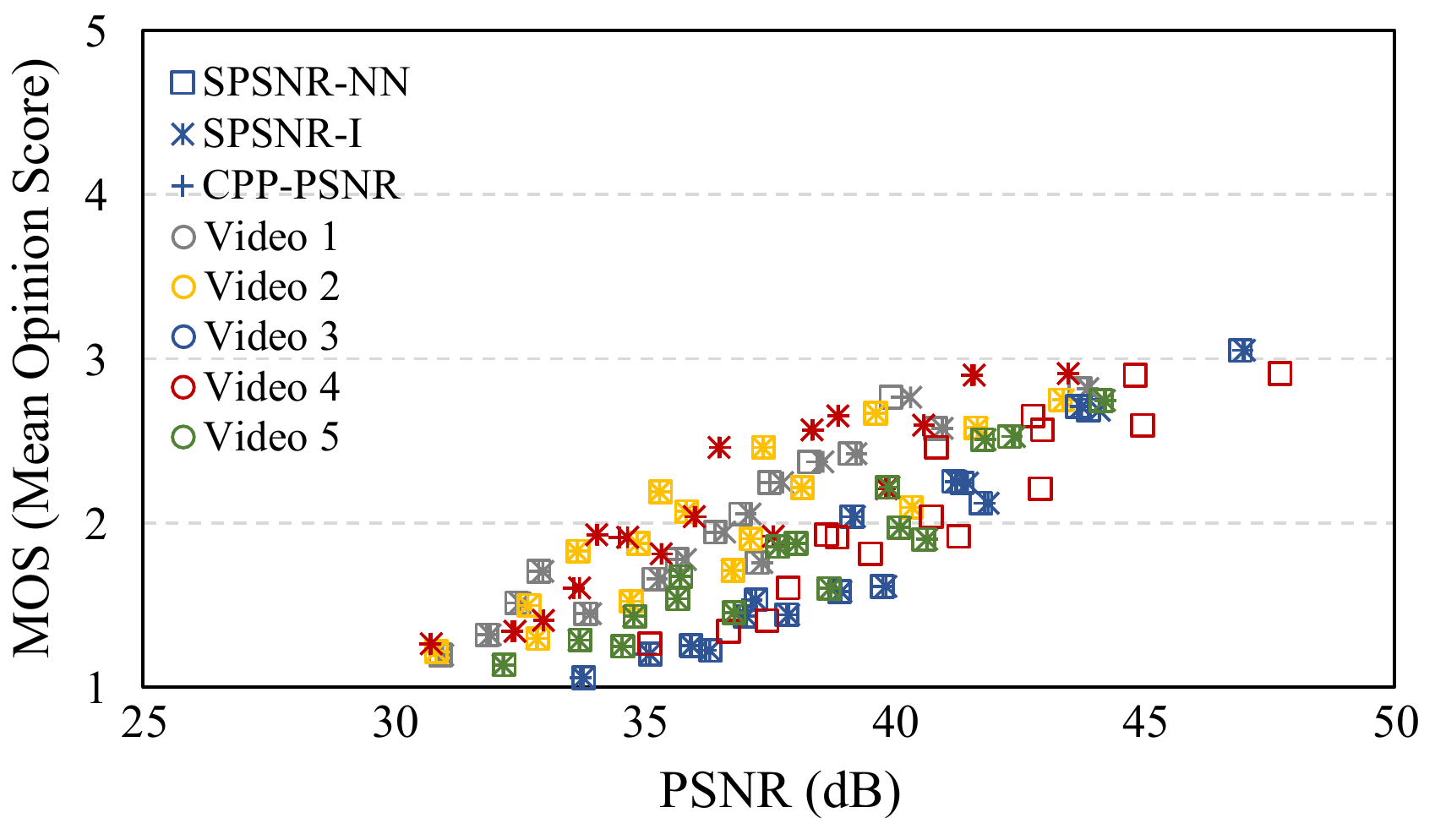}
        \caption{Cross-Format (CF) PSNR}
    \end{subfigure}
    \begin{subfigure}{0.30\linewidth}
    \includegraphics[width=\linewidth]{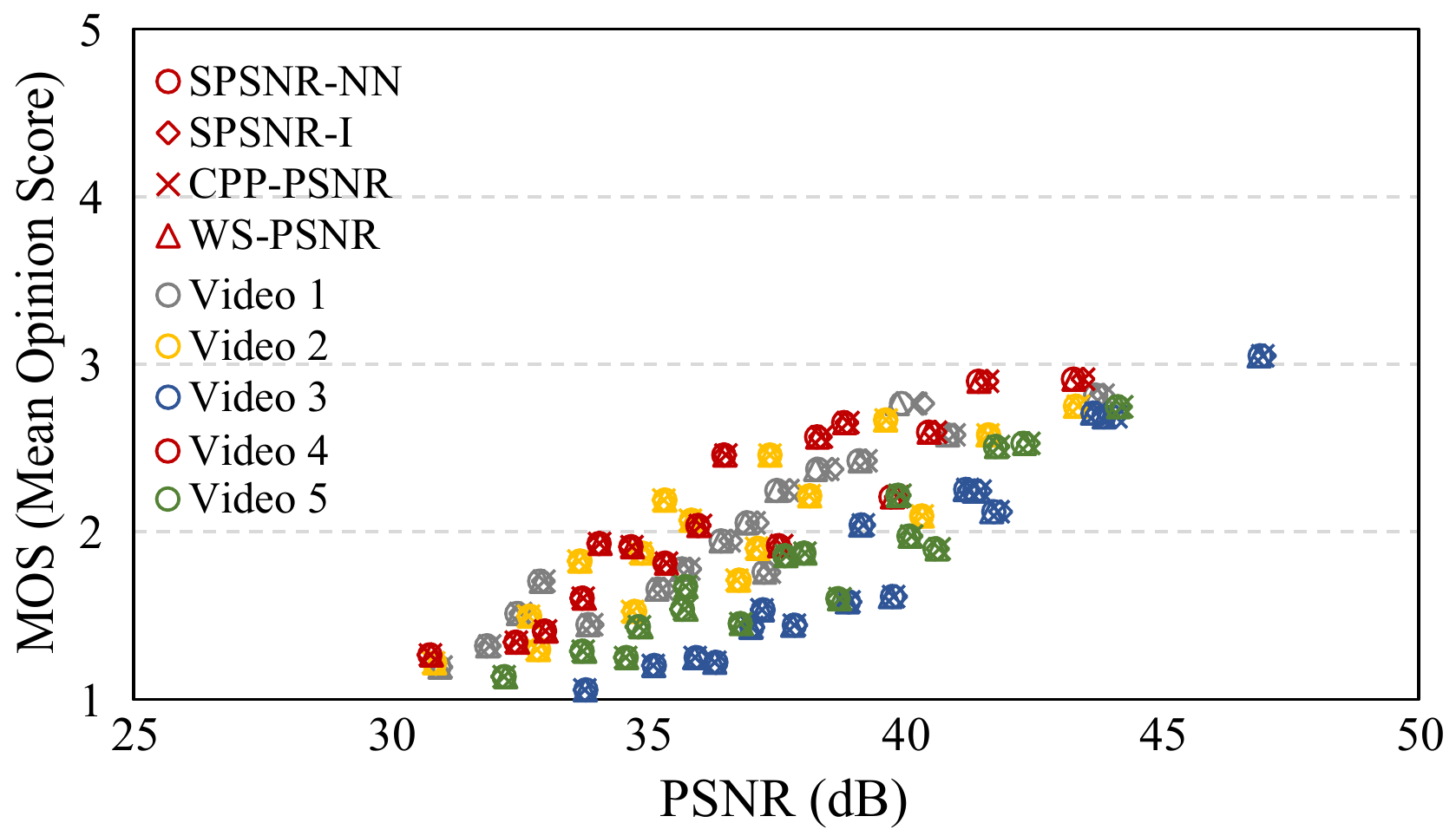}
    \caption{End-to-End (EE) PSNR}
    \end{subfigure}
    \begin{subfigure}{0.360\linewidth}
        \includegraphics[width=\linewidth]{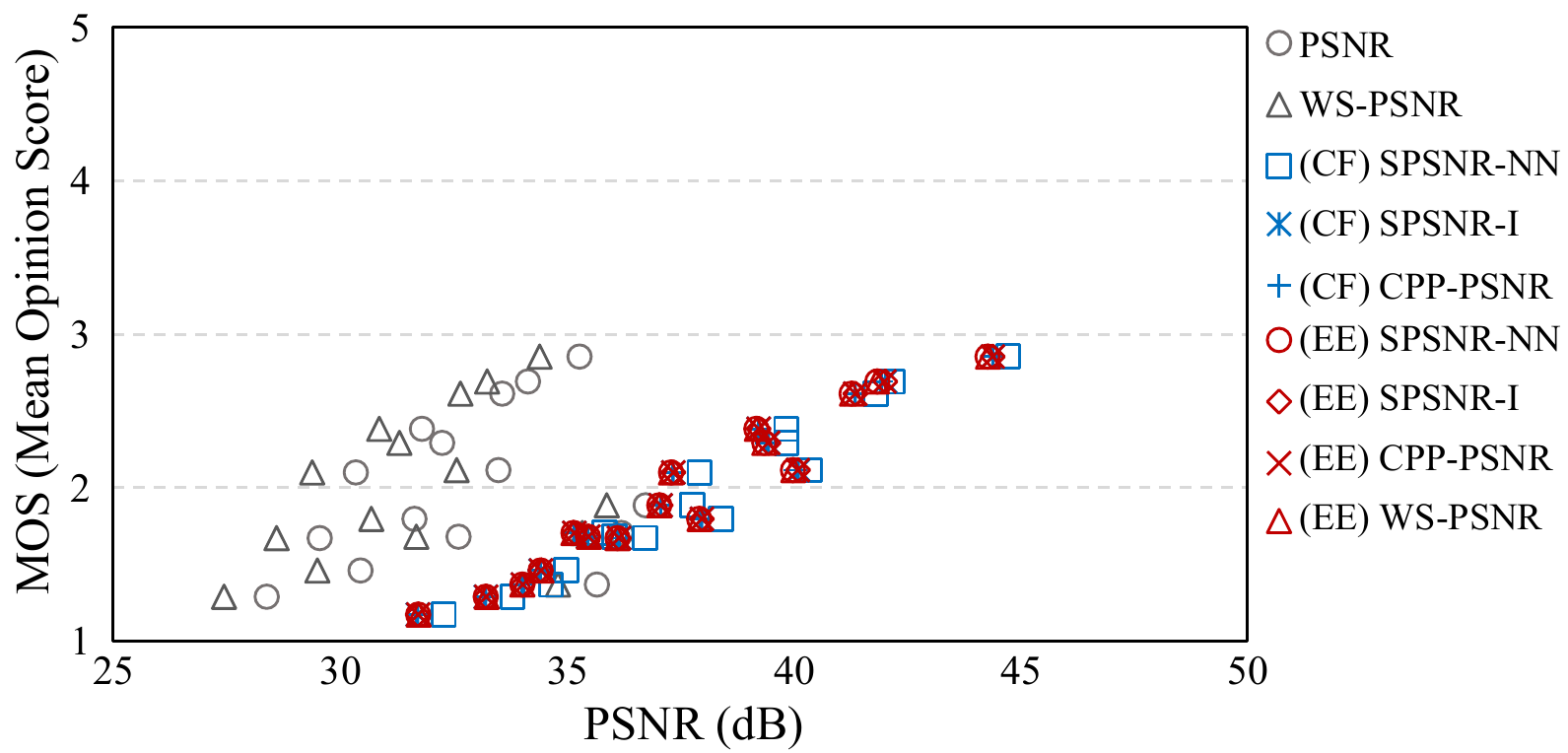}
        \caption{Average Codec, CF and EE PSNR }
        \end{subfigure}
    \caption{MOS vs PSNR related metrics of 360 video used in the experiment.}
\end{figure*}

Subjective experiments were performed in the FORCE Technology SenseLab's listening room which fulfills the requirements of EBU 3276 and ITU-R BS.1116-3\cite{ITURBS1116-3}. The test sequences were evaluated separately in terms of audio quality, video quality, and audiovisual quality test in this order. Prior to the test, the sound level (Leq) of all clips were calibrated and set to 65-70 dB (depend on the clips) at the sweet spot (listeners' head position) 1.2 meters above the floor and at a normal angle to the center-ceiling loudspeaker. We used Samsung Odyssey+ head mounted display (HMD) which has a 1440x1600 display resolution per eye, 110$^{\circ}$ horizontal field of view and 90Hz refresh rate. The HMD was operated within Windows Mixed Reality front-end platform connected to our SenseLabOnline system providing an interactive user interface within the VR video \cite{noauthor_senselabonline_nodate}. A single stimulus Absolute Category Rating (ACR) scale was used and customized with a continuous quality scale (CQS) without anchor and reference. CQS-ACR is introduced here, motivated from single stimulus rating found in SAMVIQ (Subjective Assessment Methodology for Video Quality) and the CQS found in MUSHRA and ITU-T P.800 \cite{ITURBS1534-3, rec1788bt, ITU-TP.800}.

A full factorial design was applied for the audio and video tests only. Due to the large number of audiovisual conditions, a full factorial audiovisual test was not feasible. Instead an optimal custom design of experiment (DoE) was employed to yield manageable trial size to run audiovisual quality test. A coordinate-exchange and D-optimal algorithm \cite{john1975d, meyer1995coordinate} was simulated by using DesignExpert 12 \cite{noauthor_design-expert_nodate} due to the goal is to find factors important to the process. \smallskip



\begin{figure*}[h]
\centering
\begin{subfigure}{0.3525\linewidth}
    \includegraphics[width=\linewidth]{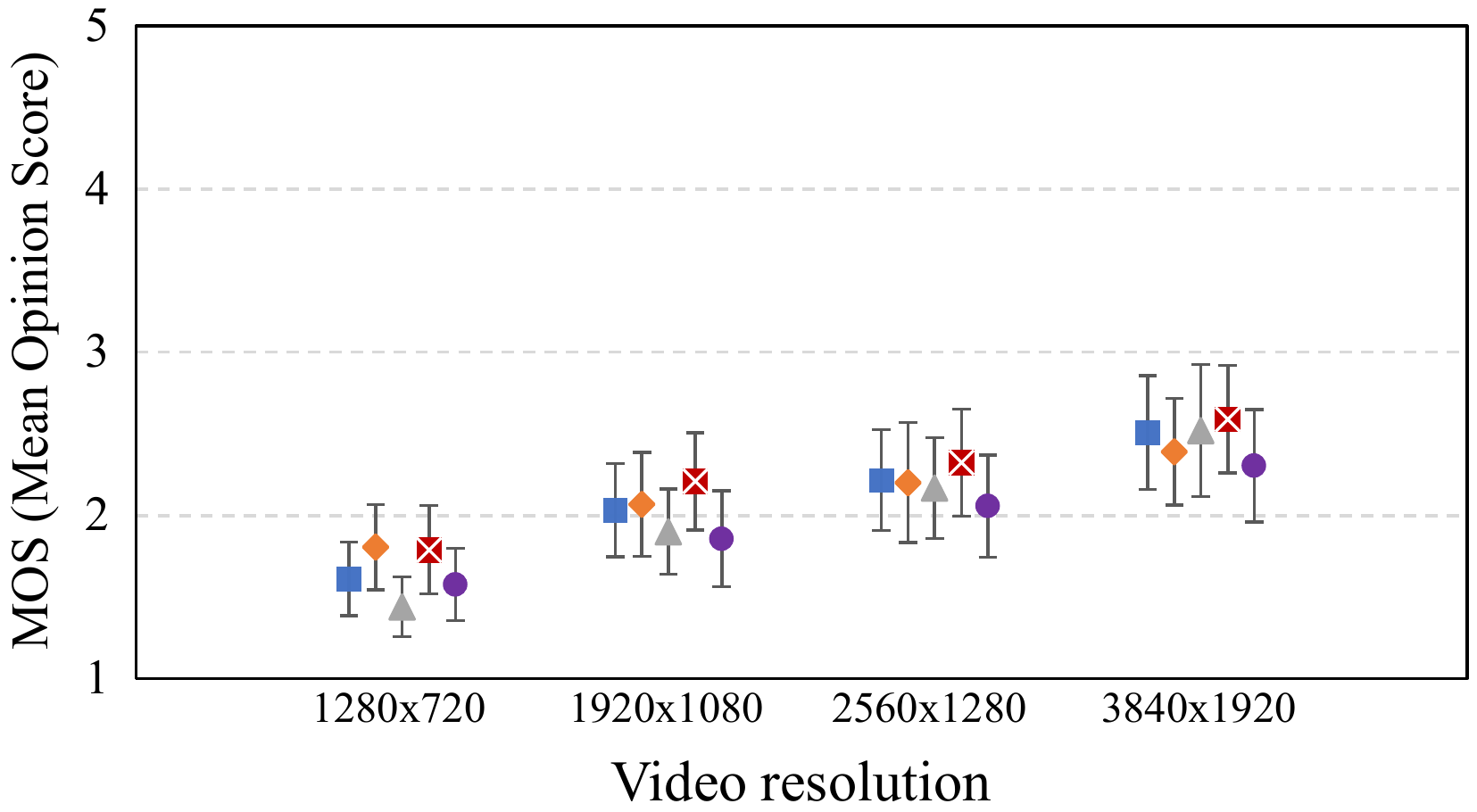}
\end{subfigure}
\begin{subfigure}{0.34\linewidth}
    \includegraphics[width=\linewidth]{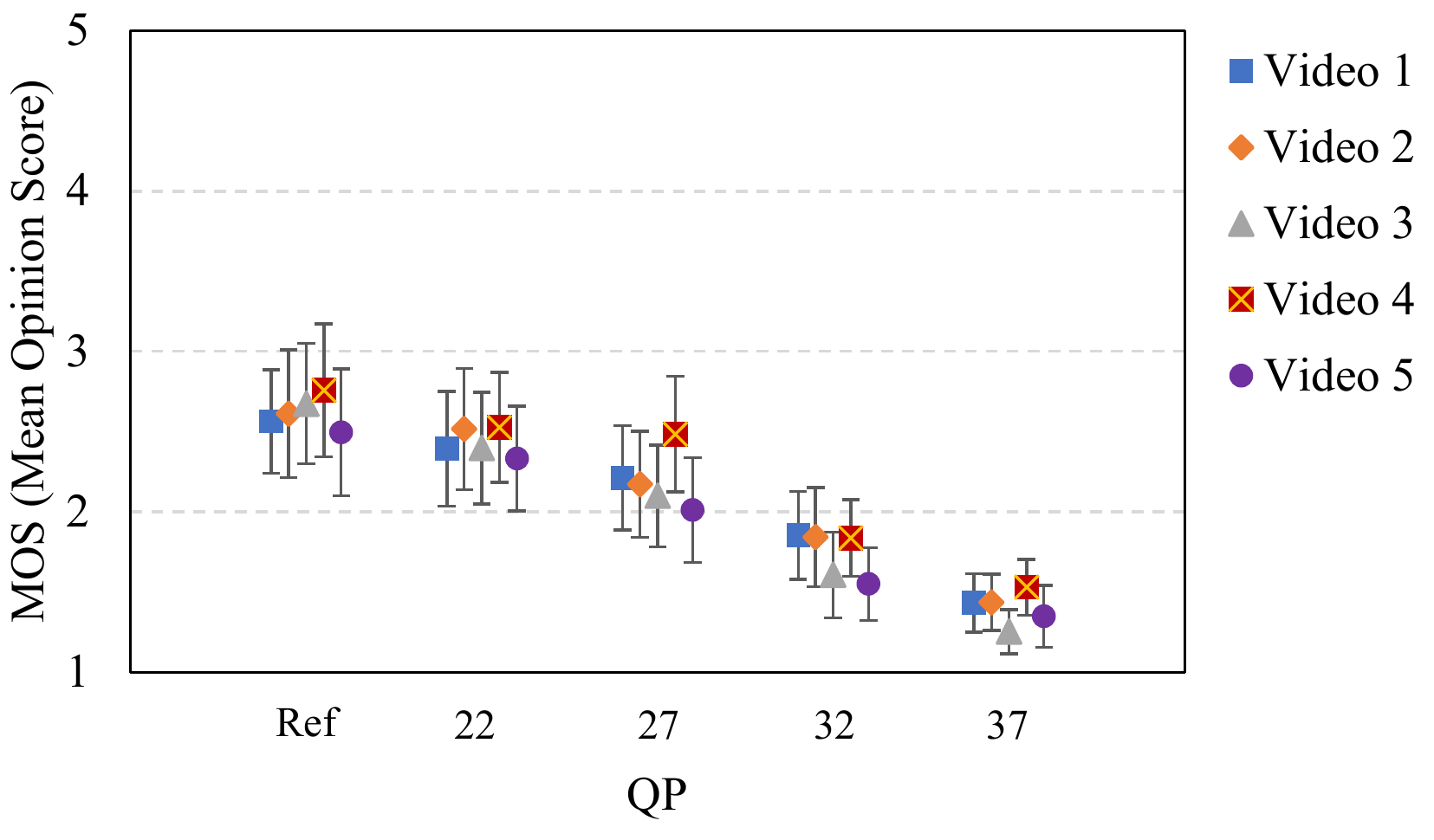}
\end{subfigure}
\caption{Perceptual video quality ($MOS_V$) vs (a) video resolution and (b) QP.}
\end{figure*}

\begin{figure*}[!h]
\centering
\begin{subfigure}{0.34\linewidth}
    \includegraphics[width=\linewidth]{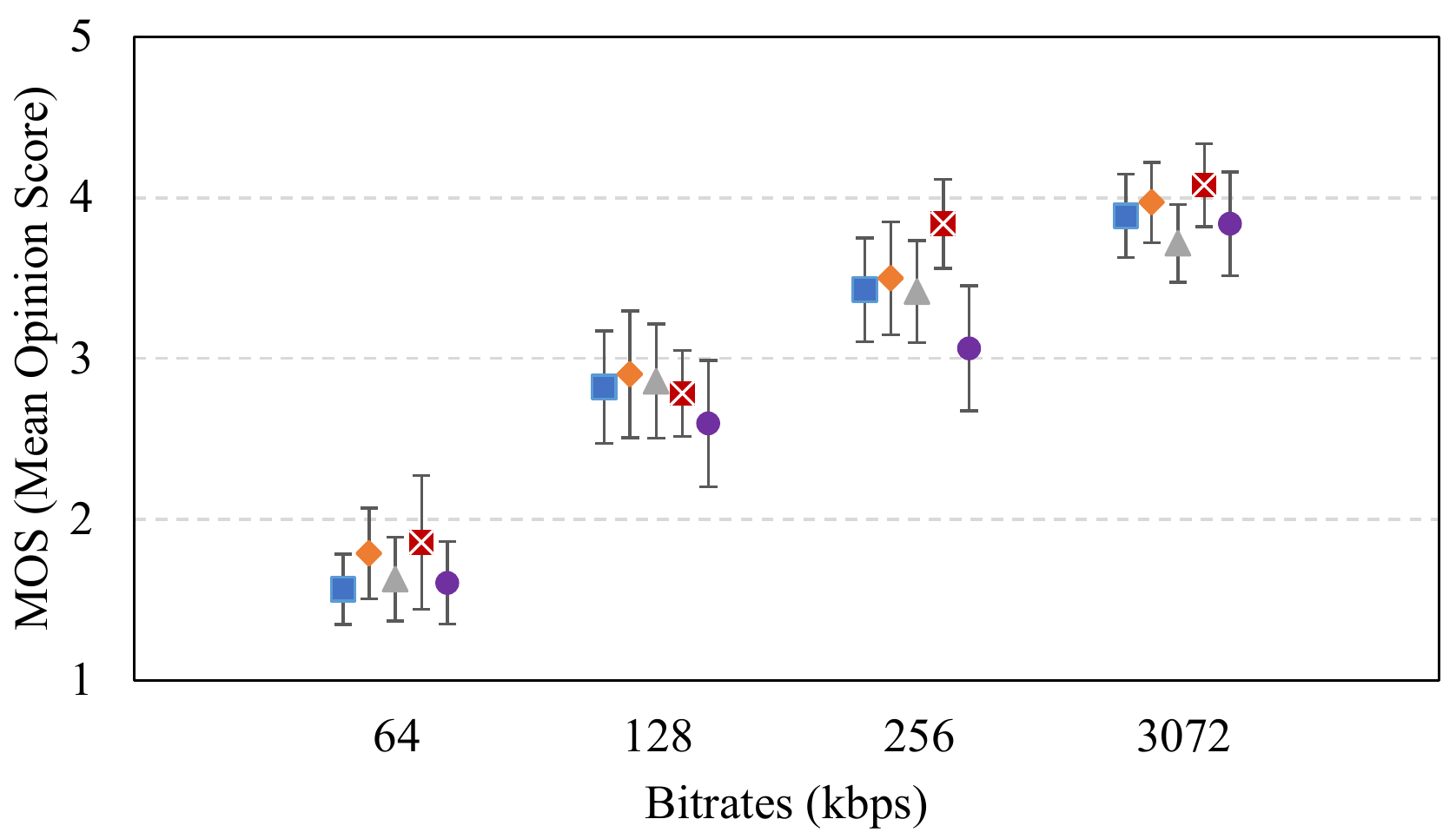}
\end{subfigure}
\begin{subfigure}{0.34\linewidth}
    \includegraphics[width=\linewidth]{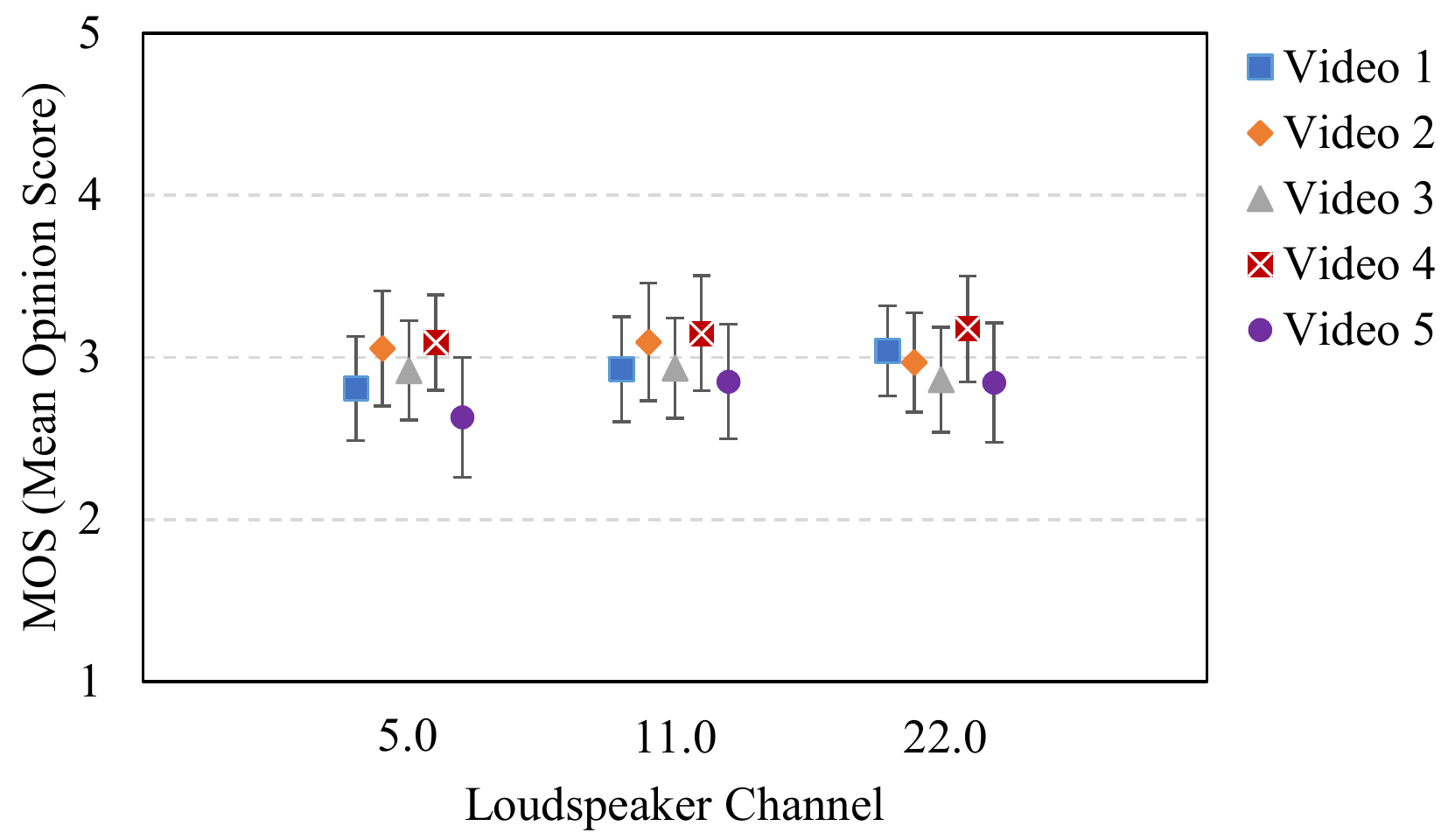}
\end{subfigure}
\caption{Perceptual audio quality ($MOS_A$) vs (a) bitrates and (b) loudspeaker channel.}
\end{figure*}
\smallskip
\subsection{Procedures}
Twenty pre-screened consumers (15 males, 5 females, mean age: 35.4, SD: 7.8) participated in this paid 4-hour study with sessions split across two days. Auditory and visual screening tests were performed prior to the experiment as close as possible according to ITU-T P.910 recommendation \cite{itup910}. Snellen chart and Ishihara plates were used to confirm visual acuity and normal color vision. Sixteen assessors were audiometrically normal and 4 self-reported normal hearing. 

During the test, the assessors were required to provide ratings using customized modular buttons which controlled 

\begin{table}[!hb]
\caption{Pearson's and Spearman's correlation between objective video metrics and $MOS_V$.}
\centering
\begin{tabular}{|c|c|c|c|c|}
\hline
\textbf Phase & \textbf Metrics & \textbf PCC & \textbf SROCC & \textbf RMSE \\
\hline

Codec			& PSNR    		& 0.6592			& 0.6557			& 0.3774\\ 			\cline{2-5} 
				& WS-PSNR		& 0.6793			& 0.6725	 		& 0.3670\\ 			\hline
Cross-Format 	& SPSNR-NN    	& \textbf{0.8384}	& \textbf{0.8311}	& \textbf {0.2642}\\ 	\cline{2-5} 
				& SPSNR-I       & 0.8218    		& 0.8105       		& 0.2847\\ 			\cline{2-5} 
				& CPP-PSNR      & 0.8215			& 0.8091        		& 0.2840\\ 			\hline
End-to-End 		& SPSNR-NN      & 0.8181			& 0.8076       		& 0.2861\\ 			\cline{2-5} 
				& SPSNR-I       & 0.8220			& 0.8113       		& 0.2847\\ 			\cline{2-5} 
				& CPP-PSNR      & 0.8219			& 0.8104       		& 0.2839\\ 			\cline{2-5} 
				& WS-PSNR       & 0.8185			& 0.8073       		& 0.2850\\ 			\hline
\end{tabular}
\end{table}

\noindent the user interface and rating scale. The rating interface was projected onto a screen for audio quality test and directly in the virtual video for video and audiovisual quality test. The tests were presented in double blind random order. The system automatically encouraged the assessors to take a short break every twenty minutes.

\section{Results}
\subsection{Objective Quality of Testing Video}
Fig. 2 illustrates the video codec PSNR values and video bitrates of the QP setting encoding parameter. All test videos except video 3 have PSNR values ranging from 30dB to 45dB. It can be seen that PSNR of video 3 starts higher than 35dB and could reach nearly 50dB in QP 22 whereas in Video 1 and Video 5 started below 35dB. The plots clearly show that video resolution and QP have significant contribution to video bitrate. Although the resolutions differ, PSNR is still nearly identical at the same QP values. All of the PSNR values have a similar trend over the bitrates except for 1280x720 resolution with a slightly different line. Moreover, video in 3840x1920 resolution has higher bitrates relatively to the lower resolution.

In total, there are 9 different PSNR metrics corresponding with different computation in 360 video processing chain \cite{boyce2017jvet}. There are five basic types of objective quality metrics including PSNR, Weighted to spherically uniform PSNR (WS-PSNR) \cite{sun2017weighted, sun2016ahg8}, spherical PSNR based on nearest neighbor position (S-PSNR-NN)\cite{he2016ahg8}, spherical PSNR with interpolation (S-PSNR-I) \cite{yu2015framework} and PSNR in Crasters parabolic projection (CPP-PSNR)\cite{zakharchenko2016ahg8}. As depicted in Fig. 3 (a-b), almost all PSNR variants in Cross-Format and End-to-End process have identical values in each video. The differences between PSNR metrics for a video coded with the same encoding parameters are relatively small. By averaging the values across all videos, Fig. 3 (c) shows that both Cross-Format and End-to-End manage to capture the influence of resolution, which the codec PSNR clearly does not as expected. 

The correlation, including Pearson Correlation Coefficient and Spearman Rank-order Correlation Coefficient and its Root Mean Square Error (RMSE), between PSNR related metrics and mean opinion subjective scores are given in Table II. 
As can be seen in Table II, $MOS_V$ shows a rather low correlation and high RMSE with PSNR measured in coding distortion (PSNR and WS-PSNR) with the PCC and SROCC value less than 0.7 and RMSE $>$ 0.36. Meanwhile, the correlations in Cross-Format and End-to-End phases are rather high ($>$ 0.80). In this study, Cross-Format SPSNR-NN produce the best results. However, SPSNR based nearest-neighbor computation depends highly on the content features where the error could increase if the features of video content is more complex. Thus, it should be noted that all videos used in this study are in the static camera mode and of only simple-moderate complexity.

\subsection{Perceptual Video Quality}
Fig. 4 shows the results for perceived video quality $MOS_V$ over the resolution of video and quantization parameter (QP). On the left side of Fig. 4, $MOS_V$ increases as the video resolution is increased but decreases when QP is increased. This is because QP regulates how much spatial detail is preserved and the QP value represents a step size on the Discrete Cosine Transform (DCT) in frequency domain. Therefore, small values of QP more accurately approximate the block's spatial frequency spectrum.

For these videos, $MOS_V$ lies within a narrow range with a maximum score of only 3 (Fair) for the best quality video presented. It is also found that the confidence interval (CI) is relatively small either in the lowest resolution or QP which means there is common agreement that the quality is very poor and highly noticeable. Furthermore, although the impact of video encoding parameters on video quality score can be concluded, based on the confidence interval (CI), only small differences are noticeable. We argue this result is due to the absence of reference content with excellent quality, which, had it been available, allowed for the quality to span the complete $MOS_V$ score range and improve the results. 

\subsection{Perceptual Audio Quality}
The mean opinion score ($MOS_A$) of perceived audio quality is presented in Fig. 5 for (a) audio bitrate and (b) loudspeaker channel. It points out that audio bitrate has a positive correlation with $MOS_A$. The highest score is obtained on the original clips (3,072 kbps). However, even the original stimuli did not yield a maximum score. There is no difference between audio clip for each bitrate and there is no dominant stimuli with the highest score in all bitrates. Statistically significant differences can be found between the audio in 64 kbps and 128 kbps, and slightly difference to 256 kbps. Only few samples have significant difference between 256 kbps and original audio.

Fig. 5 (b) shows that there is no relationship between loudspeaker channel to subjective score as seen with the overlapping confidence intervals. 
Whilst the quality and nature of the audiovisual content are well suited to this study, it would be desireable to have a larger and more critical and high quality database for future studies.
A non critical sample leads to low sensitivity of spatial changes and therefore assessors are not able to discriminate different number of loudspeaker channels. Similar finds have been found in other studies \cite{zacharov2016next} with broadcast quality programme material. Furthermore, this finding is also inline with a study from \cite{rudzki2019perceptual} which described that a lossy compression for ambisonics has a negative effect to timbral distortion thus reduce the localization accuracy. The study revealed that localization error occurs not only in \nth{1} order but also in \nth{3} and \nth{5} order ambisonics. Moreover, it is found that there is no significant median score difference for timbral distortion between ambisonic orders and bitrates. 
It would appear that in complex tasks where multiple characteristics are to be considered simultaneously by assessors, that certain characteristics dominate. In this study, audio bitrate has a clearly significant impact on the perceived sound quality, which may also be masking the small difference between number of loudspeaker channels. Therefore, in order to enhance assessors' sensitivity in doing such complex tasks thus improve the obtained results, a comprehensive training with multiple parameters could be considered prior to the test. Thereafter, further investigation might study the performance of test methods with and without a reference signal.


\subsection{Audiovisual Quality Model}
Here we presented initial investigation towards audiovisual quality model of 360 video with ambisonic audio from subjective data. The correlation of subjective data between audiovisual quality ($MOS_{AV}$) and audio quality ($MOS_A$), video quality ($MOS_V$) and the multiplication ($MOS_A.MOS_V$) were evaluated as shown in Table III. The correlation coefficients vary for each interaction and each video. Video 3 generally has the highest correlation between $MOS_A$, $MOS_V$, $MOS_A.MOS_V$ to $MOS_{AV}$. Overall the $MOS_A.MOS_V$ shows the highest correlation with $MOS_{AV}$ in all videos.

The audiovisual quality can be modelled based on linear combination of audio and video quality and the interaction. Here we evaluated six models as proposed in previous studies and has been summarized in \cite{akhtar2017audio}. The model consist of two or three predictors as shown in (1-4) and function parameter (5-6). The last two models (5-6) were proposed by \cite{martinez2014full} called weighted Minkowski and power model. 

{\footnotesize
{\begin{equation}
    MOS_{AV1} = \alpha_0+\alpha_1MOS_A+\alpha_2MOS_V+\alpha_3MOS_AMOS_V
\end{equation}
\begin{equation}
    MOS_{AV2} = \alpha_0+\alpha_1MOS_AMOS_V
\end{equation}
\begin{equation}
    MOS_{AV3} = \alpha_0+\alpha_1MOS_V+\alpha_2MOS_AMOS_V
\end{equation}
\begin{equation}
    MOS_{AV4} = \alpha_0+\alpha_1MOS_A+\alpha_2MOS_V
\end{equation}
\begin{equation}
    MOS_{AV5} =(\alpha_1MOS_A^{P}+\alpha_2MOS_V^{P})^{1/P}
\end{equation}
\begin{equation}
    MOS_{AV6} =\alpha_0+\alpha_1MOS_A^{P_1} MOS_V^{P_2}
\end{equation}
}}

\noindent where \(\alpha_0\), \(\alpha_1\), \(\alpha_2\) and \(\alpha_3\) are weighting parameters and depending on the application, they may vary between studies (\(\alpha_0\) only improves the fit of residuals and irrelevant to correlation).

\begin{table}[!h]
\caption{Correlation between audio and video quality, and their multiplication with audiovisual quality.}
\centering
\begin{tabular}{|c|c|c|c|c|c|c|c|}
\hline
\textbf{$MOS_{AV1}$}   & \multicolumn{6}{|c|}{\textbf{AV content}} \\
\cline{2-7}
\textbf{}   & \textbf{All}  & \textbf{1} & \textbf{2} & \textbf{3} & \textbf{4} & \textbf{5} \\
\hline
$MOS_A$    & 0.58 & 0.67    & 0.58    & 0.50    & \textbf{0.59}    & 0.53    \\
\hline
$MOS_V$    & 0.69 & 0.62    & \textbf {0.76}    & \textbf {0.76}    & 0.63    & 0.64    \\
\hline
$MOS_A.MOS_V$ & 0.89 & 0.90    & 0.92    & \textbf {0.93}    & 0.86    & 0.85 \\
\hline
\end{tabular}
\end{table}

\begin{table}[!h]
\caption{Accuracy between predicted and test data of audiovisual models.}
\centering
\begin{tabular}{|c|c|c|c|}
\hline
\textbf Model equation  & \textbf {PCC} & \textbf {SROCC} & \textbf {RMSE} \\
\hline
$MOS_{AV1}$ & 0.928 & 0.931	& 0.204\\ 	\hline 
$MOS_{AV2}$ & 0.919 & 0.924	& 0.216\\ 	\hline
$MOS_{AV3}$	& 0.927 & 0.934 & 0.205\\ 	\hline
$MOS_{AV4}$ & 0.917 & 0.920 & 0.215\\   \hline
$MOS_{AV5}$ & 0.916 & 0.916 & 0.218\\ 	\hline
$MOS_{AV6}$ & \textbf{0.930} & \textbf{0.935} & \textbf{0.198}\\ 	\hline
\end{tabular}
\end{table}

The subjective models were computed with 80:20 data ratio between training and test data. Pearson and Spearman correlations and RMSE were calculated as shown in Table IV. From Table IV, all models show a good fit to the data with the correlation $>$0.90. Although the power model in (6) has the highest correlation (PCC 0.930, SROCC 0.935) and lowest RMSE (0.198), the difference is relatively small. However, this result is consistent with earlier studies \cite{martinez2014full} in which it was shown that the power model had the best fit across the tested models. It should be noted that instead of using the original value of weighted parameters from referred models as in \cite{martinez2014full}, we concern to employ the model form only and also compute the weighting parameters for each model for this immersive application.

\section{Conclusion}
We carried out a subjective experiment on audio, video, and audiovisual quality with 360 video displayed on an HMD with low-bitrate ambisonic based loudspeaker reproduction, evaluated using a CQS-ACR methodology. 
The main findings can be summarized as follows:
\begin{itemize}
    \item Besides the common relationship between video PSNR, encoding parameters and subjective scores, we show that the Cross-Format and End-to-End PSNR could predict the performance across the resolutions and show a linear relationship. In perceived quality, significant differences are noticeable. However, $MOS_V$ considerably has a narrow range and a maximum $MOS_V$ is less than 3 as an impact of quality limitation as objectively shown in Fig. 2. This limited quality suggests the urgent needs of synchronized 360 video with ambisonic for audiovisual research. 
    \item There is a significant difference of $MOS_A$ across audio bitrates. Different number of loudspeaker channels show insignificant perceptual effect. Auditory stimuli (e.g ambisonic order), rating methods, task complexity and assessors' sensitivity could be the reason that this occuring.
    \item The correlation between subjective scores shows that multiplicative $MOS_{A.V}$ perform a very high correlation among the others. According to the model, although a power model has the best accuracy, a difference between AV models are indistinguishable. However, the results imply that video quality is dominant over the others. This is consistent in Internet Protocol Television \cite{garcia2009impairment} and high motion video\cite{hands2004basic} application. A proposed DoE shows good performance indicating its potential use for further investigation.
\end{itemize}

\section{Future Works}
In order to improve the performance of audiovisual quality model for immersive content, the development of transparent, broader, high quality, and critical audiovisual database is important.  
Regarding the test methodology, no standard approach exists to assess audiovisual quality for immersive content currently. The tested methodology employed in this study was able to resolve many important perceptual characteristics, but also highlighted limitations to be overcome. 

The use of reference and anchor methods such as MUltiple Stimuli with Hidden Reference and Anchor (MUSHRA) in audio and Subjective Assessment Methodology for Video Quality (SAMVIQ) in video might provide new directions for our research. 
Furthermore, a study of objective and subjective measures in immersive audiovisual content could reflect the relationship between those metrics and help in finding the model with accurate prediction. However, this study mainly lies on the subjective models. The future works will consider the objective models using the objective spatial audio metric AMBIQUAL \cite{narbutt2018ambiqual}. A number of machine learning approaches of multimodal fusion offer the opportunity to learn and propose more accurate and fast prediction.

\section*{Acknowledgment}
This research was supported by the European Union’s Horizon 2020 research and innovation programme under the Marie Skłodowska-Curie grant agreement No.765911 RealVision. We thank Prof. Angelo Farina for allowing to use his datasets. The authors also thank colleagues from FORCE Technology SenseLab who provided insight and expertise that greatly assisted the research.

\bibliographystyle{IEEEtran}
\bibliography{References}

\end{document}